\documentclass{article}
\usepackage{amsmath}
\usepackage{graphicx}

\input{tcilatex}

\begin{document}

\begin{center}
{\LARGE EPR\ Paradox , Locality and Completeness of Quantum Theory.}

\textbf{Marian Kupczynski\medskip }

\ Department \ of Mathematics and Statistics , Ottawa University

585 King Edward,Ottawa ON K1N 6N5 ,Canada

E-mail: mkupczyn@uottawa.ca\medskip 
\end{center}

\begin{quotation}
\textbf{Abstract.} The quantum theory (QT) and new stochastic approaches
have no deterministic prediction for a single measurement or for a single
time -series of events observed for a trapped ion, electron or any other
individual physical system. The predictions of QT being of probabilistic
character apply to the statistical distribution of the results obtained in
various experiments. The Copenhagen interpretation of QT\ acknowledged the
abstract and statistical character of the predictions of QT but at the same
time claimed that a state vector $\Psi $ provided complete description of
each individual physical system. The assumption that a state vector was
assigned to an individual physical system together with the postulate of \
its instantaneous reduction in the measurements was shown by Einstein,
Podolsky and Rosen to lead to so called \ EPR\ paradox for the experiments
with entangled pairs of particles. EPR concluded that a state vector could
not provide a complete description of individual systems and the question
arose whether the probabilistic predictions of QT could be derived from some
more fundamental spatio-temporal deterministic description of invisible
sub-phenomena by introduction of \ supplementary parameters. The
experimental violation of the Bell inequalities (BI) in the spin
polarization correlation experiments (SPCE) which were the implementations
of Bohm and EPR gedanken experiments, eliminated so called local and
realistic models of the sub-phenomena. Often the violation of BI has been
incorrectly interpreted as a proof of the completeness of QT or as the
violation of the locality and causality in the micro-world. In this paper we
show that local and realistic models overlooked the fact that an
experimental outcome is only the information about a particular
system-system or system -instrument interaction. Quantum phenomena are
described in terms of probabilities. It is well known that the probability
distribution is not an attribute of a dice but it is a characteristic of a
whole random experiment : '' rolling a dice''. \ Therefore quantum
probabilities are ''contextual'' because they describe the lack of knowledge
of the outcomes of experiments in contrast to the lack of knowledge of some
attributive properties of individual physical systems. We recall that the
existence of long range correlations between two random variables X and Y is
not a proof of any causal relation between these variables. Moreover any
probabilistic model used to described a random experiment \ is consistent
only with a specific protocol telling how the random experiment has to be
performed. The probabilistic model used to prove BI implied a protocol
completely inappropriate and impossible to implement for SPCE . Therefore we
conclude that the important question whether QT is predictably complete is
still open and we show how the unconventional analysis of the existing data
could help to answer it. The correct understanding of statistical and
contextual character of QT is essential for the research in the domain of
quantum information and \ quantum computing.


\end{quotation}

\subsection{Introduction}

\noindent It has been shown many years ago [1-6] that the main prerequisites
used to prove the Bell inequalities (BI) [7,8], or Clauser,Horn
,Shimony,Holt (CHSH) inequalities [9] \ such as the use of a common
probability space, joint probability distributions for non-commuting
observables etc. were inappropriate for the description of the spin
polarization correlation experiments (SPCE). Therefore the violation of
BI-CHSH\ in these experiments [10,11] gave neither information about the
nonlocality of QT nor about its completeness.

In spite of this, testing of BI-CHSH continued and several loopholes were
indicated which could explain the apparent,but not existing, violation of BI
-CHSH by the imperfection of the experimental set ups [12]. Precise new
experiments [13,14] permitted to close several loopholes, confirmed QT
predictions for the correlation functions and even allowed to detect strange
anomalies in various detection rates reported recently by Adenier and
Khrennikov [15]. Instead of looking for new loopholes it would be more
interesting, in our opinion, to use the existing data in order to find
anomalies similar to those reported in [15] or to search for some fine
structure in the data with help of various purity tests proposed by us
several years ago [4,16,17]. Only the discovery of the fine structure in the
data, not accounted for by QT , would provide a decisive proof that the
statistical description provided by QT\ of these data is incomplete ending
the debate started by EPR [18] over 75 years ago.

The limitations and inapplicability of BI , CHSH and GHZ inequalities [19]
to SPCE\ \ have been demonstrated now in numerous publications e.g.
[20-27,46-50]. Several local models [2,3,28,29,41] were able to reproduce
QT\ predictions. BI-CHSH are also violated in macroscopic experiments
discussed by Aerts [30,31] and in computer experiments of Accardi and Regoli
[21]. The strong arguments in favor of statistical and contextual character
of spin projections were given by Allahverdyan, Balian and Newenhuizen [32].
Instead of rejoicing that there was no contradiction between QT and locality
[25] \ many members of physics community continue to marvel at the picture
of two perfect random dices giving completely correlated outcomes etc.

One may wonder why it is so ? \ One reason could be that they do not
understand the implications of the use of a common probability space and
joint probability distributions for the protocol of a random experiment .
Perhaps the authors criticizing BI were using too technical or/and too
condensed language e.g.:'' The main hypothesis needed to prove Bell-type
inequalities is the assumption that the probabilities estimated in various
SPCE can be calculated from one sample space ( probability space) by
conditionalization.'' [5] .

The other reason could be that after Harry Potter and other science fiction
stories everything seems to be possible and magic explanations for the
correlations are much more attractive than those based on a common cause or
a common sense.

Similar reasons may explain perhaps the fact known for years and explained
recently in detail by Ballentine [34] that :''Once acquired, the habit of
considering an individual particle to have its own wave function is hard to
break.... though it has been demonstrated strictly incorrect,''.

Fortunately we are at the fourth Vaxj\"{o} conference and in my opinion
slowly a consensus is starting to build up around statistical contextual
interpretation (SCI) of QT [20,27,33-35]. According to this interpretation
the information gathered in the measurements in all different experimental
contexts provides the only reliable contextual information about a ''state''
of the identically prepared physical systems whose properties are measured.

It is not an easy process because during the conference several discussions
showed that people are attributing different meaning to the words such as :
probability,\ contextual, observables, measurement, photon and local realism
and they have their own mental images of invisible sub-phenomena underlying
every physical observable phenomenon or experiment.

In this paper we will give additional arguments in favor of SCI. The paper
is organized as follows. In the section \ 2 we define the meaning which we
attach to the terms and notions such as: phenomena, sub-phenomena, physical
reality, filters, contextual, observables, probabilities etc. In the section
3 we recall shortly the EPR paradox and the explanation given by SCI. In the
section 4 we explain why according to SCI there is no strict
anti-correlations of ''measured'' spin projections for each ''individual EPR
pair'' in the singlet state. In the section 5 we criticize in historical
perspective various proofs of BI-CHCH including the most recent one given by
Larsson and Gill [36] and we explain why the prerequisites used in these
proofs are not valid in SPCE.\ In the section 6 we discuss how the
predictable completeness of QT could be tested.

\subsection{IMPORTANT NOTIONS}

The main assumption in physics is that there exists a material world
(physical reality) in which we can observe various phenomena and with which
we may interact by performing various repeatable experiments. \ Another
assumption is the existence of physical laws which are responsible for the
richness of phenomena and for the regularities observed in our experiments.
We have no doubt that the Moon exists when we do not look at it but of
course when we look at it we can perceive it in different colors, in
different phases etc. We do not have intuitive images of electrons and
photons but we have the abstract mathematical model given by QT\ able to
describe quantitatively various phenomena they are participating in. The
importance of the interplay of ontic and epistemic realities in QT was
recently discussed in some detail by Atmaspacher and Primas [37] and Emch
[38].

The mathematical models are always of limited validity and apply to
particular phenomena. To describe the motion of the Moon around the Earth we
use a model of a material point following well defined trajectory obtained
by solving Newton's differential equations of motion. To explain its phases
the Moon is modeled as a sphere. To describe the details of the formation of
a crater on the Moon, when a meteorite hits it, we need much more detailed
and complicated mathematical model. This is why we will probably discover
one day that at very short distances the extendedness of the hadrons and of
other elementary particles plays more important role [39] than in the
standard model we are using today.

Therefore the statement that a given theory for example QT provides the most
complete description of the individual physical systems lacks humility since
as Bohr said [40]: ''The main point to realize \ is that the knowledge
presents itself within a conceptual framework adapted to account for
previous experience and that any such frame may prove too narrow to
comprehend new experiences.''

\subsubsection{Phenomena and Sub-Phenomena}

Phenomena \ produce observable effects. Sub-phenomena are invisible. For
example we can see a track left by a single charged high energy elementary
particle on the picture from the bubble chamber. To describe its trajectory
we can use a model of a point-like particle moving according to the laws of
classical physics but of course there is an underlying microscopic invisible
sub- phenomenon leading to the track formation and if we wanted to explain
it in detail we should go beyond the classical electrodynamics. If this high
energy elementary particle collides with the proton from a hydrogen atom in
a bubble chamber then we see many out- going tracks from the collision point
what is a new phenomenon : a creation of several new particles during the
collision. To describe this phenomenon we have to go beyond the quantum
electrodynamics. We have to prepare in an accelerator a collimated beam of
''identical elementary particles'' and to observe several collision events
in the bubble chamber. The only reproducible regularities, besides the
conservation laws, we assumed \ to be valid, are of statistical nature and
can be predicted by some abstract mathematical model \ providing the
probabilities for different possible outcomes of the collision. The
intuitive description of invisible sub-phenomena taking place during the
collision is not provided.

The phenomena described by QT\ are all of this nature. Any experimental
set-up can be divided into three parts: a source preparing the ensemble of
''identical'' physical systems , an interaction/filtering part and the
detectors/counters part which produces time- series of observable events :
clicks on various detectors, dots on a screen, tracks in a bubble chamber
etc. One may have an impression that nowadays we are able to perform the
experiments on the individual physical systems such as an electron or an ion
in some trap. However in order to find any statistical regularity in the
data from these experiments we have to reset initial conditions in the trap
and repeat these experiments several times obtaining again the ensemble of
measurements performed on the ''identical'' physical systems. For the
extensive discussion of general experimental set-ups see [42]. QT gives the
probabilistic predictions for the distribution of outcomes \ obtained in
various phenomena without providing intuitive models of invisible
sub-phenomena. One encounters paradoxes only if incorrect models are used to
describe these sub-phenomena . A source of light is not a gun and photons
are not small bullets etc.

\subsubsection{Attributes and Contextual Properties}

An attributive property is a constant property of an object which can be
observed and measured at any time and which is not modified by the
measurement e.g.: inertial mass, electric charge etc. A contextual property
is a property revealed only in specific experiments and characterizes the
interaction of the object with the measuring apparatus. Let us mention
Accardi's chameleon which is green on a leaf and brown on a bark of a tree
[22,25].

In classical physics we assume that measurements of various attributive
properties possessed by an individual physical system are compatible what
means that they can be measured simultaneously\ or in any sequence giving
always the same results. In quantum physics the contextual properties are
known after the measurements e.g. as a click on a detector placed behind a
polarization filter. The measurements of incompatible properties cannot be
performed simultaneously and the measurement of one of them destroys the
information about the other one. Various sequences of these measurements
lead to various probability distributions of the outcomes [42]. The
measurements of attributive properties are called by Accardi [25] passive
dynamical systems and those of contextual properties active dynamical
systems.

In QT contextual properties of individual systems are of statistical
character because they may be only deduced from the properties of pure
ensembles they are members of [41] ''...a value of a physical observable,
here a spin projection, associated with a pure quantum ensemble and in this
way with an individual physical system, being its member, is not an
attribute of the system revealed be a measuring apparatus; it turns out to
be a characteristic of this ensemble created by its interaction with the
measuring device. In other words the QM is a contextual theory in which the
values of the observables assigned to a physical system have only meaning in
a context of a particular physical experiment''.

Another argument in favor of \ SCI of QT comes from probability theory. The
probabilities are only the ''properties'' of random experiments [5]:\ ''
talking about the probabilities we should always indicate the random
experiment needed to estimate their values''. QT provides the algorithms to
find various probabilities therefore it is a contextual theory. The
contextuality in this sense is the fully objective property of the Nature.
Even if nobody observes the collisions of high energy protons they are
described by the same probability distributions of the possible outcomes no
matter where it happens.

The probabilities found in QT\ do not describe one particular random
experiment but a whole class of equivalent random experiments which are
assumed to be repeatable as many times as needed.

\subsubsection{Probabilities, Correlations and Causality}

It is not obvious how to define the probability, what is the randomness etc.
These important topics were discussed \ recently in detail by Khrennikov in
his stimulating book [20].

We illustrate here these difficulties by two simple examples, the more
detailed discussion may be found in [20,22,42]

1) Let us consider a random experiment which can give only two outcomes: 1
or -1. We repeat this experiment 2n times and we obtain a time series of the
results:1,-1,1,-1,...,1,-1. By increasing the value of n the relative
frequency of getting 1 can approach 1/2 as close as we wish suggesting that
the probability of getting 1 in each experiment is equal to 1/2. Of course
it is incorrect because if we analyze the time series of outcomes in detail
\ we see that we have a succession of the couples of two deterministic
experiments or one deterministic experiment keeping memory of the previous
result which is called a periodic two dimensional Markov Chain.

2) Let us toss now a fair coin assigning 1 for ''head'' and -1 for ''
tail''. We get again a time series: 1,1,-1,-1,-1,1,-1..with relative
frequencies which tend to 1/2. There is no apparent structure in this series
so the hypothesis of the independent and identical repetitions of the same
experiment seems to be satisfied and we say that a complete description of
this experiment is provided by a single number :a probability of getting 1
which is equal to 1/2. Of course we believe that if we knew all parameters
describing the sub-phenomena of tossing experiment we could predict each
individual result using the laws of the classical physics. The randomness is
here only due to the lack of control of these parameters.

The statisticians and probabilists invented many tests in order to test the
randomness and independence in such time series but conclusions from any
statistical study is valid only on a given level of confidence. Moreover the
series \ formed by the consecutive decimals of the number $\pi$ passed all
the tests of the randomness in spite of the fact that any consecutive
decimal is strictly determined. Without any completely conclusive measure of
randomness we have to limit ourselves for any practical applications to
generators of pseudo-random numbers which passed the known tests of the
randomness.

One of the postulates of Copenhagen interpretation was that in a measurement
process a measured value of a physical observable is chosen among all
possible values of this observable with a given probability and in a
completely random way. It was also believed that this indeterministic
behavior of quantum ensembles could not be explained by the lack of control
of some hidden variables describing deterministic interactions of individual
members with measuring devices. This intrinsically indeterministic behavior
of individual quantum ''particles'' was believed to provide a new standard
of the randomness which could serve to produce the unbreakable keys in
quantum cryptography.

The fact that the strong correlations created by the source in SPCE\ survive
the filtration and measurement processes is a strong argument against purely
random behavior of individual systems during these processes. The individual
systems have to carry a memory of their preparation coded in some parameters
and a measuring device described by its own uncontrollable microscopic
parameters has to act in some deterministic way to produce an observable
outcome without destroying completely the memory how the systems were
prepared at the source. It is well known that a strong correlation between
two random variables has nothing to do with a causal relation between these
variables. For example the average price of oil in a given year is
correlated with the average salary of Anglican priests in the same year due
to the common cause which is inflation. The existence of strong correlations
between non-interacting physical systems which interacted in the past was
analyzed for the first time by EPR [18] and led them to conclude that the
description of these phenomena provided by QT\ is incomplete. In the next
section we discuss shortly their paper.

\subsection{EPR-PARADOX}

EPR consider two systems I and II which are permitted to interact from t=0
to t=T and which evolve freely and independently afterwards. The state of \
I or II for t$\geq$T can be found only by the reduction of wave packet. Let a%
$_{1}$, a$_{2}$, a$_{3}$.. be the eigenvalues of \noindent some physical
observable A to be measured on the system I and u$_{1}$,u$_{2}$,u$_{3}$.. a
complete set of corresponding orthogonal eigenfunctions . At the moment of
measurement T$_{1}$ $\geq$ T of the observable A on the system I the wave
function $\Psi(x_{1},x_{2})$ of the system I+II is given by 
\begin{equation}
\Psi(x_{1},x_{2})=\underset{n}{\sum}\psi_{_{n}}(x_{_{2}})u_{n}(x_{1})  
\tag{(1)}
\end{equation}
If the measurement of A gives a$_{k}$ then the wave function is reduced to c$%
_{k}\psi_{_{k}}(x_{_{2}})$ $u_{k}(x_{1})$ \ \ where $\psi_{_{k}}(x_{_{2}})$
is, up to normalization constant, the wave function of the system II
immediately after the measurement of A on the system I has been completed
and the result a$_{k}$ known.

If instead of A we decided at t=T$_{1}$ to measure on I another
non-commuting observable B with the eigenvalues b$_{1}$,b$_{2}$,b$_{3}$ and
a complete set of orthogonal eigenfunctions v$_{1}$,v$_{2}$,v$_{3}$... then
instead of the formula (1) we would have

\begin{equation}
\Psi(x_{1},x_{2})=\underset{s}{\sum}\varphi_{_{s}}(x_{_{2}})v_{s}(x_{1})  
\tag{(2)}
\end{equation}
If b$_{r}$ was obtained then the wave of the system II immediately after the
measurement of B on the system I would have been , up to normalization
constant, $\varphi_{_{s}}(x_{_{2}}).$

In their paper EPR conclude :'' Thus it is possible to assign two different
wave functions ( in our example $\psi_{_{k}}(x_{_{2}})$ $and$ $%
\varphi_{_{s}}(x_{_{2}})$ ) to the same reality (the second system after the
interaction with the first'' . In each case the functions are assigned with
certainty and without disturbing the system II. If one assumes that the wave
functions are in one to one correspondence with the states of individual
physical systems one obtains a contradiction called EPR Paradox.

Of course according to SCI there is no paradox because the wave functions \ $%
\psi _{_{k}}(x_{_{2}})$ $and$ $\varphi _{_{s}}(x_{_{2}})$ \ describe only
different sub-ensembles of the ensemble of the particles II. The eigenvalue
expansions (1) and (2) being mathematical identity describe different
incompatible experiments.\ They imply the existence of the long-range
correlations between the measurements performed on the non-interacting
separated physical systems . The state $\Psi (x_{1},x_{2})$ is not
factorized and is an example of the entangled state so popular nowadays. The
discussion following the EPR\ paper was recently reviewed in detail in
[27].\ Some arguments of EPR\ were rejected but nobody was able to prove
that QT provided the complete description of individual physical systems.

\subsection{SINGLET STATE}

The most studied example of the entangled state is a singlet spin state.
Spin version of the EPR experiment was proposed by Bohm [43]. Using SCI\ we
analyze here only the predictions of QT for SPCE. The detailed discussion of
EPR-B paradox in a spirit of SCI may be found in [27].

The singlet spin state vector for the a system of two particles has the
form: 
\begin{equation}
\Psi_{0}=\left( \mid+\text{ }\rangle\otimes\text{ }\mid-\text{ }\rangle\text{
- }\mid-\text{ }\rangle\otimes\text{ }\mid+\text{ }\rangle\right) \sqrt{1/2} 
\tag{(3)}
\end{equation}

where the single particle vectors $\mid+$ $\rangle$ and $\mid-$ $\rangle$
denote ''spin up'' and ''spin down '' with respect to some common coordinate
system.

If we ''measure'' the spin of the particle \#1 along the unit direction
vector \textbf{a} and the spin of the particle \#2 along the unit direction%
\textbf{\ \ }vector \textbf{b} , the results will be correlated and for the
singlet state the correlations are described by the correlation function: 
\begin{equation}
E(\mathbf{a},\mathbf{b})=\left\langle \Psi_{0}\left| \sigma_{\mathbf{a}\text{
\ \ }}\otimes\text{ }\sigma_{\mathbf{b}\text{ \ \ }}\right| \Psi
_{0}\right\rangle =-cos\text{ }\theta_{\mathbf{ab}}\ \   \tag{(4)}
\end{equation}

where $\ \sigma_{\mathbf{a}\text{ \ \ }}=$ \textbf{\ }$\mathbf{\sigma}\bullet
$ \textbf{a} \ \ and $\sigma_{\mathbf{b}\text{ \ \ }}=$ \textbf{\ }$\mathbf{%
\sigma}\bullet$ \textbf{b} \ \ denote the components of the Pauli spin
operator in the directions of the unit vector \textbf{a} and \textbf{b}
respectively and $\theta_{\mathbf{ab}}$ is the angle between the directions 
\textbf{a} and\textbf{\ b}. Since $E(\mathbf{a},\mathbf{b})=-1$ for $\
\theta_{\mathbf{ab}}$ =0 it was concluded that the results of the spin
projection measurements for each individual couple of particles are strictly
anti-correlated. It was pointed out in [41] that this conclusion is
unjustified since according to SCI the state vector $\Psi_{0}$ allows only
to find statistical distribution of outcomes without giving a deterministic
prediction for any individual outcome. The reason is that sharp directions
and angles do not exist in the Nature. Fuzzy measurements in QT\ have been
studied for years and many important results have been obtained \ [44,45].

Each spin polarization correlation experiment (A,B) is defined by two
macroscopic orientation vectors\ \textbf{A} \ and\textbf{\ B} being some
average orientation vectors of the analyzers [41,26,27]. More precisely the
analyzer \textit{\ A} is defined by a probability distribution d$\rho _{A}(%
\mathbf{a})$ , where \textbf{a }are microscopic direction vectors \textbf{a}$%
\in O_{A}$ $\ $and $O_{A}=\left\{ \mathbf{a}\in S^{(2)};\left| 1-\mathbf{a}%
\cdot\mathbf{A}\right| \leq\varepsilon_{A}\right\} .$ Similarly the analyzer
B is defined by its probability distribution d$\rho _{B}(\mathbf{b}).$ \
Therefore even if the detectors and filters were perfect , no detection
loophole, the idealized QT prediction for the correlation function $E(%
\mathbf{A},\mathbf{B})$ would be given not by the formula (4) but by a
smeared formula:

\ 
\begin{equation}
E(\mathbf{A},\mathbf{B})=\underset{O_{A}}{\int}\underset{}{\underset{O_{B}}{%
\int}-cos\text{ }\theta_{\mathbf{ab}}\ }d\rho_{A}(\mathbf{a})d\rho _{B}(%
\mathbf{b})\ \   \tag{(5)}
\end{equation}

The quantitative effect of smearing of $cos$ $\theta_{\mathbf{ab}}$ in the
formula (5) can be very small but $\ E(\mathbf{A},\mathbf{A})\neq-1$ and
there are no strict anti-correlations between measured polarization
projections. Of course in SPCE \ the formulas (4) and (5) have to include
additional factors to account for the efficiencies of detectors, various
transmission coefficients etc. The formula (5) and similar formulas for
joint probabilities of detection [41,26,27,32] confirm only the fundamental
contextuality of QT due to which a spin projection on a given axis is not a
predetermined attribute of an individual physical system recorded by a
measuring device but it is created in the interaction of the system with
this device. As we already told the correlations between far away
measurements suggest the existence of supplementary parameters keeping the
memory of the preparation stage and describing the invisible sub-phenomena
during the measurement process..

\subsection{BELL INEQUALITIES}

Let us describe a typical SPCE\ in the language of observed phenomena.

A pulse from a laser hitting a non-linear crystal produces two correlated
physical fields propagating with constant velocities towards the far away
detectors. Each of these fields has a property that it produces clicks when
hitting the photon- detector. We place two polarization analyzers A and B in
front of the detectors on both sides and after interaction of the fields
with the analyzers we obtain two correlated time- series of clicks on the
detectors. Each analyzer is characterized by its macroscopic direction
vectors, which may be changed at any time. By changing the direction vectors
we have various coincidence experiments labeled by (A, B ) where \textbf{A}
and \textbf{B} are the macroscopic direction vectors for the analyzers A and
B respectively.

In QT the crystal is described as a source of couples of photons in a spin
singlet state and the ensemble of these couples is described approximately
by the state vector (3). It is well known that the individual photons are
neither localizable nor visible and they do not behave as point-like
particles following some classical trajectories. Nevertheless the mental
picture of correlated photon pairs travelling across the experimental set-up
and carrying their own unknown spins (intrinsic magnetic moments ) whose
projections on any direction are predetermined by a source and recognized by
the polarization analyzers is commonly used in the discussions of SPCE.
Knowing that such description of the sub-phenomena is inaccurate Bell tried
to formulate the most abstract probabilistic local hidden variable model in
order to explain the spin polarization correlations in a singlet state
predicted by QT. As we told above in the experiment (A,B) two time -series
of outcomes are produced with each outcome being 1 or -1. The main
assumptions in so called local realistic hidden variable model (LRHV)
proposed by Bell [7] are:

1. Individual outcomes are produced locally by corresponding analyzers A and
B.

2. There are some uncontrollable hidden variables $\lambda\in\Lambda$
determining the value of individual outcomes. In the experiment (A,B) the
outcomes are obtained as values of some bi-valued functions on $\Lambda$
such that A($\lambda,\mathbf{a})=\pm1$ and B($\lambda,\mathbf{b})=\pm1$
respectively where\textbf{\ a} and \textbf{b} denote the settings of the
analyzers (we keep here for purpose the original Bell notation \textbf{a}
and \textbf{b} instead of \textbf{A} and\textbf{\ B} used above).

3 .The probability space $\Lambda$ and the probability distribution $\
\rho(\lambda)$ do not depend on \textbf{a }and\textbf{\ b.}

Since the probability distribution of hidden variables is prepared at the
source far away from the detectors Bell wrongly believed that the assumption
3 is another consequence of the locality. The assumptions 1-3 allow to write
the correlation function E(a,b) as:

\ 
\begin{equation}
E(\mathbf{a},\mathbf{b})=\underset{\Lambda}{\int}A(\lambda,\mathbf{a}%
)B(\lambda,\mathbf{b})\rho(\lambda)d\lambda\ \ \   \tag{(6)}
\end{equation}

Using the formula (6) for any couple of directions of analyzers, BI and CHSH
inequalities below can be proven. \bigskip\ 
\begin{equation}
\bigskip\left| E(\mathbf{a},\mathbf{b})-E(\mathbf{a},\mathbf{b}^{\prime
})\right| +\left| E(\mathbf{a}^{\prime},\mathbf{b}^{\prime})+E(\mathbf{a}%
^{\prime},\mathbf{b})\right| \leq2\ \ \   \tag{(7)}
\end{equation}

In 1976 we met John Bell in Geneva, and we left him few handwritten pages
with our comments concerning the limitations of his proof. In particular we
pointed out that if the hidden variables $\lambda$ describing each pair of
''particles'' were couples of bi-valued, strictly correlated, spin functions
S$_{1}$ and S$_{2}$ on a sphere such that measured outcomes for each pair
were the values of S$_{1}$(\textbf{a}) and S$_{2}$(\textbf{b}) then one
could not use the integration over the set of all of these functions as it
is done in the formula (6). Moreover we indicated that in this case one
should try to prove BI by using the estimates of the correlation functions :
the empirical averages obtained by averaging the sums of the products S$_{1}$%
(\textbf{a})S$_{2}$(\textbf{b}) over all pairs in long runs of the
corresponding experiments. If E(\textbf{a,b}) is replaced by its estimate
the proof of (7) may never be rigorous because the error bars have to be
included and one has also to assume that the sets of spin functions
describing the couples in the runs from different experiments are exactly
the same what is highly improbable due to the richness of the uncountable
set of spin functions on a sphere. These ideas in their final more mature
form were only published later in a series of papers [4,5,16,41].

In the meantime Pitovsky [2,3] constructed the spin functions on the sphere
for which the integral (6) could not be defined and proposed a local hidden
variable model based on these functions able to reproduce the predictions of
QT and violating BI. Aerts[6,30] inspired by Accardi's paper[1]\ showed that
BI can also be violated in macroscopic experiments. De Baere[46,47] pointed
out that BI might be violated due to the non-reproducibility of a set of
hidden variables.

The Pitovsky model was difficult to understand. \ We simplified it and
rendered fully contextual in [41]. The spin functions were strictly
correlated but due to the smearing over the microscopic directions there was
no strict anti-correlations.

In [5] we gave several arguments why BI could not be proven rigorously using
the empirical averages and we showed that the use of the unique probability
space implied the experimental protocol which was incompatible with SPCE. In
conclusion we wrote:''The various SPCE cannot be replaced by one random
experiment of the type discussed above and in our opinion this is the reason
why the Bell inequalities do not hold. The various probabilities appearing
in their proofs are counter-factual and have nothing to do with the measured
ones''

Let us clarify here about which one random experiment we were talking. The
k-variate random variable X=(X$_{1},...,$X$_{k})$ and joint probability
distribution on a unique probability space $\Lambda$ were invented in order
to describe the following random experiment: we take a large random sample
of members from some population we measure the values of X$_{i}$ $i=1,..,k$
on each member in a sample obtaining an individual outcome for the
experiment as a set \ of k numbers (x$_{1,..,}$x$_{k})$ .\ The empirical
joint distribution for the frequencies of these outcomes gives the
information about the joint probability distribution characterizing the
whole population. \ From this joint probability distribution one may obtain
by conditionalization the marginal probability distributions for any single
random variable X$_{i}$ or for any group of them. This is exactly the
protocol of the random experiment implied by the formula (6): pick up a pair
described by $\lambda,$ measure the spin projections for this pair in all
directions \ etc. what is of course impossible.\ We thought that the
arguments presented against BI were convincing enough to stop further
speculations concerning their violation but we were wrong. \ Probably some
papers were simply unknown or not understood. With growing interest in
quantum information and speculations about faster than light communications
in EPR experiments it was necessary to provide an up-to-date refutation of
BI and of nonlocality of QT. It was done e.g. by Accardi et al. [22,25],
Khrennikov[20], Hess and Phillip[24], Kracklauer[28] and by
myself.[22,26,27]. We already explained why the formula (6) did not apply to
SPCE .

The correct formula which may be used to describe locally the sub-phenomena
in any particular experiment SPCE for the couple of analyzers (A,B) is:

\ 
\begin{equation}
E(\mathbf{A},\mathbf{B})=\underset{\Lambda_{AB}}{\int}A(\lambda_{1},\mathbf{a%
})B(\lambda_{2},\mathbf{b})\rho(\lambda_{1},\lambda_{2})d\rho _{A}(\mathbf{a}%
)d\rho_{B}(\mathbf{b})d\lambda\ \ \   \tag{(8)}
\end{equation}

where $\Lambda_{AB}=\Lambda_{1}\times\Lambda_{2}\times\Lambda_{A}\times
\Lambda_{B}$ $\ $\ with ($\lambda_{1},\lambda_{2},\mathbf{a,b)\in}\Lambda
_{AB}$ and d$\lambda$ is a shorthand notation for the measure on $\Lambda
_{1}\times\Lambda_{2}$ \ for which the integral makes sense.

One of the most recent reformulations of CHSH theorem was given by Larsson
and Gill [36]. \ Instead of $\Lambda_{AB}$ they use a unique probability
space $\Lambda$ as in formula (6) saying\ :''...the particles travelling
from the source carry some information about what the result would be of
each possible measurement at the detectors. This information is denoted $%
\lambda$ above,and can consist of anything , from a simple ''absolute''
polarization to some complicated recipe of what each result measurement will
be, for each setting of the detector parameter. What it is exactly is not
important the very existence of such information will be referred to as ''
Realism'', the $\lambda$, in a sense is '' the element of reality'' that
determines the measurement result.''. Without writing explicitly the formula
(6) they conclude that under these prerequisites the CHSH theorem cannot be
violated and therefore the Local Realistic hidden variable models are
impossible. Next \ authors give an interesting discussion of how various
experimental factors such as visibility , efficiency etc. may prevent the
violation or prevent the application of the theorem getting a formula (9)
which we rewrite here in a simplified form:

\ 
\begin{equation}
|E(\mathbf{a},\mathbf{b|}\Lambda\mathbf{(}A\mathbf{,}B\mathbf{)})-E(\mathbf{a%
},\mathbf{b}^{\prime}|\Lambda(A,B^{\prime}))|+\left| E(\mathbf{a}^{\prime},%
\mathbf{b}^{\prime}|\Lambda(A^{\prime},B^{\prime }))+E(\mathbf{a}^{\prime},%
\mathbf{b|}\text{ }\Lambda(A^{\prime},B))\right| \leq4-2\delta\   \tag{(9)}
\end{equation}

where $\ \Lambda\mathbf{(}C\mathbf{,}D\mathbf{)\subset}$ \ $\Lambda$ denotes
a subset of $\Lambda$ describing the experiment (C,D) and $\delta$ is some
minimum probability of the overlap of the subsets used in the formula (9). \
In spite of the fact that the authors did not list very strong assumption
concerning the existence of the joint distributions on the unique
probability space $\Lambda$ in their proof of (7) and (8) we completely
agree with them that if the ''Realism '' is understood as a strict
predetermination of the experimental outcomes at the source then Local
Realistic hidden variable models are impossible. All local models able to
reproduce the QT predictions [21,41,25,28,29] and to violate the equalities
(7) are contextual as well as contextual are all SPCE. The correct formula
is (8) and there is no overlap between $\Lambda_{AB}$ and $\Lambda_{CD}$ if C%
$\neq A$ or B$\neq D.$ Therefore BI\ or CHSH cannot be proven. The only
formula which can be proven is the formula (8) with $\delta=0$ which of
course does not violate QT predictions. Using the language of Larsson and
Gill the information $\lambda$ does not determine the future measurement
results. The hidden information in the moment of the measurement ,carried by
the particles, about the preparation at the source is stored in ($%
\lambda_{1},$ $\lambda_{2}).$ The information which predetermines the
outcome of the measurement for the analyzer A is stored in ($\lambda_{1,}%
\mathbf{a})$ where \textbf{a }describes a microscopic state of analyzer A in
the moment of measurement. The information what the outcome would be is not
created at the source and decoded with mistakes by the analyzer but it is
created in the interaction with the analyzer and known only after the
measurement is completed.

The hidden variable model of underlying sub-phenomena given by (8) is
intuitive, local and contextual .\ According to Accardi's terminology [25]
such probabilistic model describes an adaptive dynamical system. Another
simple hidden variable model of SPCE\ has been proposed recently by Matzov
[29].

We see that testing of BI-CHSH cannot help us to check the completeness of
QT. We should test instead the predictable completeness of QT.

\subsection{\protect\bigskip PURITY TESTS AND PREDICTABLE COMPLETENESS}

Let us consider an experiment in which we have a stable source producing a
beam of '' identical invisible particles'' whose intensity is measured by
the clicks on some detector. When we pass this beam by some quantum filter F
we obtain a beam having different properties and reduced intensity. The
detailed discussion why quantum filters are not the selectors of preexisting
properties is given in [42]. If by repeating our experiment several times we
discover that the relative frequencies converge to some number p(F) we may
interpret it as a probability that an individual particle from the beam \
will pass the filter F. According to SCI the claim that QT gives the
complete description of the individual system being a member of some pure
quantum ensemble may be only understood in the sense that the probabilistic
predictions of QT\ provide complete description of the ensembles of the
outcomes of all possible measurements performed on this pure quantum
ensemble.

A standard interpretation of QT did recognize the importance of a pure
quantum state and defined it as a state of physical system which passed by a
maximal filter or on which a complete set of commuting observables was
measured. The immediate question was what to do if we did not have a maximal
filter or how could we know that a filter used was a maximal one?\ We found
this definition highly unsatisfactory and we analyzed in 1973 various
general experimental set-ups containing \ the sources of some hypothetical
particle beams, detectors (counters), filters, transmitters and instruments
[42].This analysis led us to the various conclusions which are pertinent to
the topic of this paper:

1) Properties of the beams depend on the properties of the devices and
vice-versa and are defined only in terms of the observed interactions
between them. For example a beam \ b is characterized by the statistical
distribution of outcomes obtained by passing \ several replicas of this beam
by all available devices d$_{i}$. A device d is defined by the statistical
distribution of the results it produces for all available beams b$_{i}$. All
observables are contextual and physical phenomena observed depend on the
richness of the beams and of the devices.

2) In different runs of the experiments we observe the beams b$_{k}$ each
characterized by its empirical probability distribution. Only if an ensemble
\ss\ of all these beams is a pure ensemble of pure beams we can associate
estimated probability distributions of the results with the beams b$\in$\ss\
and eventually with the individual particles who are forming these beams.

3) A pure ensemble \ss\ of pure beams b is characterized by such probability
distributions s(r) which remain approximately unchanged:

(i) for the new ensembles \ss$_{i}$ obtained from the ensemble \ss\ by the
application of the i-th intensity reduction procedure on each beam b$\in$\ss

(ii) for all rich sub-ensembles of \ss\ chosen in a random way

In order to test the validity of the Optical Theorem \ we decided to test
whether the initial two-hadron states prepared for the high energy collision
are mixed with respect to the impact parameter. Therefore we reviewed in a
series of papers several non-parametric statistical purity tests which could
be used [17] and together with Gajewski [51] we performed the purity tests
for $\pi^{-}d$ charge multiplicity distributions using the raw data from the
Cambridge-Cracow-Warsaw collaboration in which the deuterium filled bubble
chamber was exposed to a $\pi^{-}$ beam of momentum 21GeV/c. We wanted to
find significant differences between the data obtained in the different
accelerator runs. \ If the initial state is pure, the different channels
should be randomly distributed in time with some fixed probabilities of the
appearance. If one concentrates on the appearance of two groups of the
channels one can obtain the time ordered sequences of 0 and 1 such as :
1000110000... The randomness of these sequences can be tested in different
ways . In one of the tests the hypothesis that the distribution was random
could be rejected on the significance level:as low as of 0,0014. Since we
considered our paper mainly as the illustration of various testing methods
we did not insist on the importance of this result hoping that it will be
confirmed by others.

In 1984 we noticed that the purity tests could be used also to test
completeness of QT because :'' The main feature of any theory with
supplementary parameters is that the quantum pure ensembles become mixed
ensembles of the individual systems characterized by the different values of
these new parameters. There is a principal difference between a pure
statistical ensemble and a mixed one. The pure ensemble is homogeneous , a
mixed one should reveal a fine structure.''[16]. \ If the source is
producing a pure beam of particles all runs of the experiment should be
highly compatible. If the source is producing a mixed ensemble, the mixture
could vary slightly from one run to another. We could also hope to change
its composition by using some intensity reduction procedures. The purity
test may be defined more rigorously as follows.

Let O be a stable source of particles and $\gamma$ a measuring device of
some physical observable $\gamma$X. A set S=\{x$_{k;}$ $k=1,...,m$\}, where x%
$_{k}$ denote the measured values of $\gamma$X, is a sample drawn from some
statistical population of the random variable X associated with the
observable $\gamma$X. If b$_{i}$ is a beam of m$_{i}$ particles produced by
the source O in the time interval [t$_{i}$ , t$_{i}$ +$\Delta$t] we obtain a
sample S$_{i}$ when $\gamma$X is measured the beam b$_{i}$. By using j-th
beam intensity reduction procedure applied to the beam b$_{i}$ we obtain a
family of new beam b$_{i}$(j) , j=1,,,,n . Measuring $\gamma$X on the beams b%
$_{i}$(j) we obtain n new samples S$_{i}$(j). We state that the beams
produced by the source O are pure only if we cannot reject the hypothesis H$%
_{0}:$

H$_{0}$: All the samples S$_{i}$ and S$_{i}$(j) for different values of t$%
_{i}$ and $\Delta$t are drawn from the same unknown statistical population.

To test H$_{0}$ one has to use the statistical non-parametric compatibility
tests such as : Wilcoxon-Mann-Whitney test, normal scores test, rank or run
tests [17]. These tests can be used to analyze any existing experimental
data in particular the data from SPCE. Of course the rejection of H$_{0}$
proves only the impurity of the ensembles which were incorrectly described
in QT\ as pure ensembles but it could be also an indication that QT is not
predictably complete. It is not easy to show that QT is not predictably
complete because the mathematical language it uses is very rich and flexible
[27,42] allowing a good fit to the experimental data.

To prove the predictable incompleteness of QT we need something more. The
results of any experiment may be represented as a time series of various
possible outcomes. If there are k different outcomes possible QT describes
this time series as a sample drawn at random from some particular
multinomial probability distribution . The outcomes should appear therefore
randomly in time with given probabilities. If one could detect some temporal
fine structure in this time series or to find a stochastic model able to
explain it \ then it would mean that QT\ does not provide a complete
description of the experimental data obtained in this experiment. Several
methods are used to study and to compare empirical time-series: frequency or
harmonic analysis, periodograms etc.[52,58].

Due to the limited efficiencies of detectors and other imperfections of the
experimental set-ups one has always a dilemma whether and how one should
correct the data in order to obtain a ''fair sample'' of outcomes to be
compared with the theoretical model of the phenomenona. Adenier and
Khrennikov analyzed recently the data from SPCE\ of Greg Weihs et al.[13]
and found several interesting anomalies which survived various data
correction attempts and which were not accounted for by the current
description provided by QT. To elucidate these anomalies one would like to
have more information about the calibration tests performed by the group
before the experiment such as numbers of counts on the detectors: without
spin analyzers on one of the sides and on both sides, with coincidence
circuitry on and off, with the singlet source replaced by another source
producing two beams having known spin polarization etc.

\subsection{CONCLUSIONS}

In spite of experimental imperfections we do not believe that the violation
of BI in SPCE is due to the unfair sampling. The main reason, is that the
probabilistic models used to prove BI are not valid for SPCE\ .

We hope that the arguments presented in this paper will cut short all
speculations about the nonlocality observed in SPCE \ and will promote the
statistical contextual interpretation (SCI) of QT .

SCI is free of paradoxes and shares Einstein's conviction that the
probabilistic description of the phenomena is due to the lack of knowledge
and control of the underlying sub-phenomena. At the same time SCI agrees
with Bohr that the measured value of the physical observable is not
predetermined and that it has only meaning in a specific experimental
context which must be always included in any model aiming to describe the
sub-phenomena. SCI agrees with Bohr that QT gives only the probabilistic
predictions for the phenomena but SCI does not say that more detailed
description of the underlying sub-phenomena is impossible.

According to SCI the mysterious long range correlations in SPCE are due to
the memory of the preparation at the source preserved in the sub-phenomena.
Several local descriptions of sub-phenomena based on this idea e.g.
[2,3,41,28,29] were able to reproduce the prediction of QT. Similar long
range correlations exist in the macroscopic world. For example the violent
earthquake in the middle of the ocean causes strong correlations between
random variables such as :the force and the height of the Tsunami waves,
force of the winds, number of victims etc. on far away shores [22]. Also in
statistical physics there exist the long range correlations between
coordinates and coarse -grained velocities of the Brownian particles which
interacted in the past what was proven by \ Allahverdyan, Khrennikov and
Nieuwenhuizen [53]. In view of this any speculation how the measurement
performed on one photon influences the behavior of the other photon from the
EPR pair is completely unfounded.

A general belief was that the language of Hilbert spaces and probability
amplitudes used by QT could not be deduced from the classical theory of
probability. This belief was shown to be incorrect by Khrennikov [54] \ who
developed a probability calculus of conditional probabilities depending
explicitly on experimental contexts and was able to reconstruct as a special
case the probabilistic formalism of nonrelativistic quantum mechanics. In
another paper [55] he showed that quantum averages \TEXTsymbol{<}O%
\TEXTsymbol{>}=Tr($\widehat{O}$ $\rho)$ can be obtained as approximations of
expectation values in some prequantum classical statistical field theory.

One cannot obtain a proof of incompleteness of QT by constructing ad hoc
local hidden variable models able to reproduce some predictions of QT . The
convincing proof of incompleteness would require a construction of a general
model providing \ a consistent description of all phenomena described by QT
and able to give more detailed predictions of these phenomena than those
given by QT what seems to be a formidable task.

Testing the predictable completeness\ of QT\ seems easier and more
promising. \ The outcomes of the physical experiments can be represented by
some\ numerical time series. QT takes for granted the randomness of these
time series and gives only the probabilities of the appearance of various
outcomes. Any significant deviation from the randomness or the discovery of
some reproducible fine structure in these series, not explained by QT, would
prove that QT is not predictably complete and that a more detailed
description of the phenomena is needed.

For example to describe effectively the behavior of cold trapped ions the
continuous quantum evolution had to be supplemented by some quantum jumps
obeying some stochastic L\'{e}vy process what was demonstrated by Claude
Cohen-Tannoudji and collaborators [56]. Similarly it seems plausible that
some new description may be needed in order to describe in detail the data
from beautiful experiments with ultra slow propagation of coherent light
pulses in Bose-Einstein condensates reported recently by Lene Vestergaard
Hau\ and collaborators [57].

The discovery of new fine structures in the data would be important by
itself but also it would give additional clear argument against treating the
quantum state vectors as attributes of individual physical systems which can
be manipulated instantaneously \ Such interpretation and instantaneous
manipulations of qubits are often used in the domain of quantum computing
[59].

Bohr considered QT as a theory of quantum phenomena and insisted on the
''wholeness'' of such phenomena [40]. If we have to talk about invisible
sub-phenomena , what is inevitable, we have to use the most precise language
in order to avoid paradoxes and confusion.

\subsection{ACKNOWLEDGMENTS}

The author would like to thank Andrei Khrennikov for the warm hospitality
extended to him during this enjoyable and interesting conference.

\end{document}